\documentclass[preprint,journal]{vgtc}             

\onlineid{1242}

\preprinttext{To appear in IEEE Transactions on Visualization and Computer Graphics.}

\vgtccategory{Research}

\graphicspath{{figs/}{figures/}{pictures/}{images/}{./}} 

\usepackage{booktabs}

\usepackage{mathptmx}                  

\usepackage{amsmath}
\usepackage{colortbl}

\newcommand{\name}{DeepConnect}
\AtEndPreamble{%
  \captionsetup[figure]{name=Fig.,labelsep=period}%
  \captionsetup[table]{name=Table,labelsep=period}%
}

\title{{\name}: A Visual Analytics System for Bridging Interdisciplinary Research Collaborations}

\author{%
  Yingchaojie Feng,
  Zekai Shao,
  Yiqun Sun,
  Yixuan Tang, and
  Anthony K. H. Tung
}

\authorfooter{
  \item
    Yingchaojie Feng, Yixuan Tang, and Anthony K. H. Tung are with the National University of Singapore. E-mail: \{feng.y, dcstyx, dcstunga\}@nus.edu.sg.
  \item
    Zekai Shao is with Fudan University. E-mail: zkshao23@m.fudan.edu.cn.
  \item 
    Yiqun Sun is with Magellan Technology Research Institute. E-mail: duke.sun@mtri.co.jp.
}

\abstract{Interdisciplinary research collaboration is crucial for scientific innovation, but it remains difficult to initiate in practice. Existing collaborator discovery approaches are often constrained by disciplinary boundaries and static researcher profiles that do not reflect the specific context of a new collaboration goal. As a result, researchers struggle to translate open-ended collaboration goals into domain-specific tasks, evaluate candidate researchers' fit and complementarity, and establish common ground before initial contact. To address these challenges, we present DeepConnect, an LLM-augmented visual analytics system for interdisciplinary collaborator discovery. DeepConnect translates collaboration ideas into domain-specific tasks, retrieves relevant papers to ground cross-domain exploration, and provides coordinated visualizations for exploring and comparing candidate researchers. It further reveals terminology gaps and overlaps across domains and supports publication-grounded conversation rehearsal to help users prepare for outreach. We evaluate DeepConnect through two case studies, a user study, and a component-level evaluation, showing its value for complementary team formation, idea refinement, and pre-contact preparation. The DeepConnect website is available at \url{https://deepconnect.sg}.
}


\teaser{
  \centering
  \includegraphics[width=\linewidth, height=0.5626\linewidth]{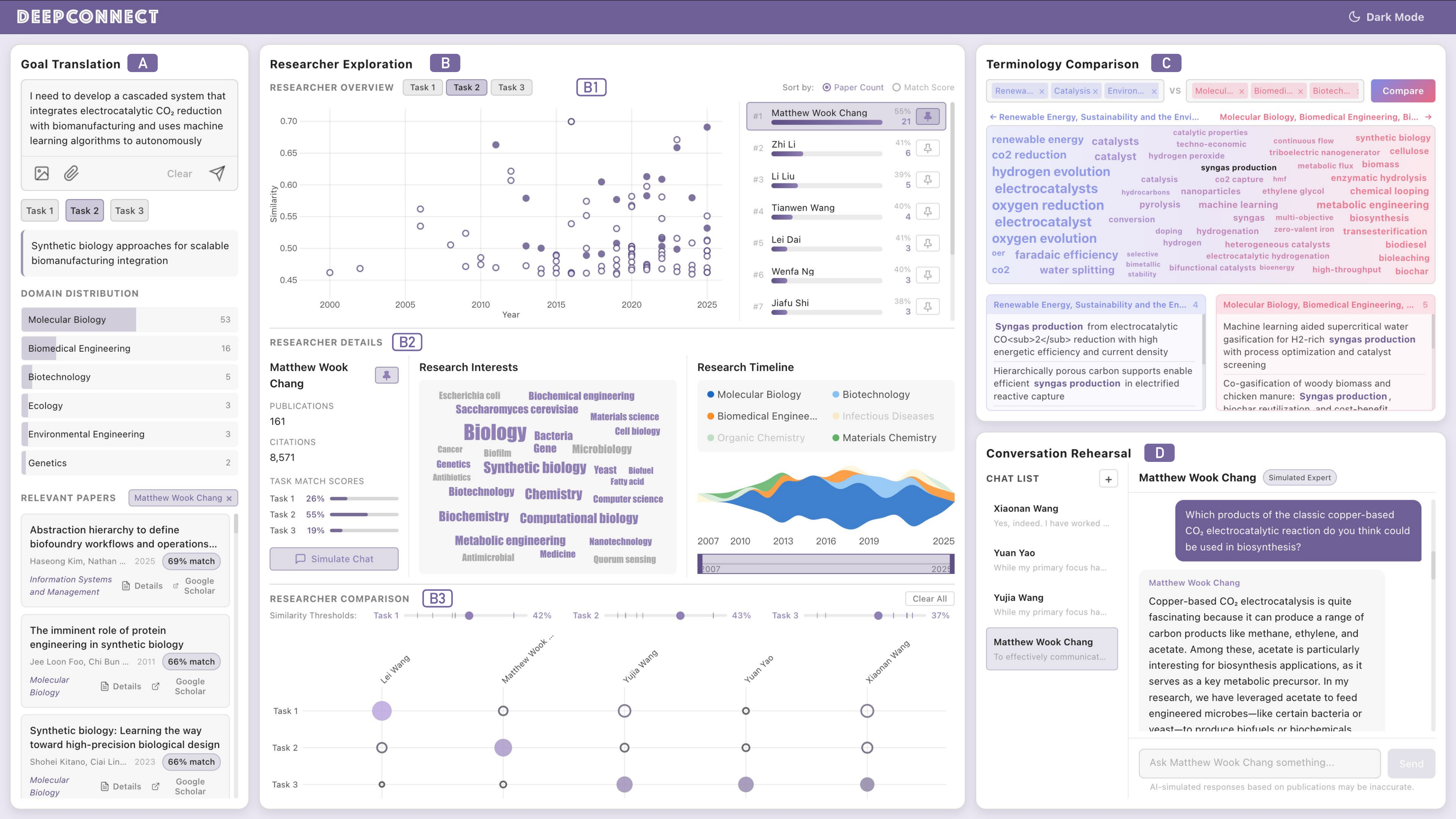}
  \captionsetup{type=figure,name=Fig.,labelsep=period}
  \caption{DeepConnect is a visual analytics system designed to bridge interdisciplinary research collaboration. The interface features multiple coordinated views: (A) The Goal Translation view decomposes overarching research goals into actionable domain-specific tasks; (B) The Researcher Exploration view enables multi-dimensional evaluation of candidate expertise and complementarity; (C) The Terminology Comparison view bridges semantic gaps by visually mapping terminology gaps and overlaps; and (D) The Conversation Rehearsal environment supports publication-grounded communication rehearsal prior to formal engagement.}
  \label{fig:system}
}

\begin{document}


\firstsection{Introduction}
\maketitle

Interdisciplinary collaboration is widely recognized as a crucial driver of scientific innovation \cite{committee2005facilitating, porter2009science, fortunato2018science, okamura2019interdisciplinarity}, particularly for complex problems that transcend traditional boundaries \cite{committee2005facilitating, okamura2019interdisciplinarity}. Integrating diverse methodologies and knowledge bases enables researchers to address complex problems intractable within a single domain \cite{stirling2007general, uzzi2013atypical}. Despite these benefits, initiating such partnerships remains practically difficult \cite{cummings2005collaborative, stokols2008science}. A promising conceptual goal does not spontaneously translate into an actionable collaboration. Instead, researchers must navigate the intricate process of identifying suitable cross-domain candidates, assessing their complementary expertise, and establishing semantic common ground prior to formal engagement.

Current collaborator discovery practices inadequately support this complex transition \cite{zhang2023scholarly, latif2018vis, beck2024puresuggest}. While serendipitous encounters and collegial introductions foster initial trust, they rely on constrained personal networks and lack scalability \cite{granovetter1973strength, burt2004structural, newman2001structure}. Academic search engines provide broader access but struggle with cross-domain exploration \cite{balog2012expertise, zhang2023scholarly, beck2024puresuggest}. Their reliance on exact keyword matching forces users to guess unfamiliar domain terminology. Furthermore, because search results are predominantly publication-centric, users must manually synthesize isolated papers to reconstruct a candidate's evolving trajectory and evaluate complementary fit. Although automated recommendation methods \cite{zhou2017collaborator, wang2020attributed, hassan2025supervised, yu2014predicting} reduce search effort, they frequently prioritize historically similar or proximate researchers, failing to account for the specific context of novel collaborative objectives \cite{mcpherson2001birds, mcnee2006being}.

Drawing on these limitations and our formative study, we identify three core challenges in interdisciplinary collaborator discovery. \textbf{(C1) Extracting Methodological Needs from Domain Contexts.} Researchers often articulate collaborative goals through the narratives and terminology of their own disciplines. This obscures the underlying methodological requirements, making it difficult to search for applicable methods and expertise from other domains \cite{mao2019data, uzzi2013atypical}. \textbf{(C2) Synthesizing Expertise from Fragmented Literature.} Assessing a candidate requires a comprehensive understanding of their research trajectory. Since standard search engines retrieve isolated publications, users struggle to manually aggregate these records to evaluate collaborative fit \cite{luo2023impact}. \textbf{(C3) Bridging Semantic Gaps in Communication.} Even when promising candidates are identified, differing academic paradigms complicate initial interactions. Researchers need support to overcome these communication barriers and assess project feasibility before initiating formal outreach \cite{caccamo2023boundary}.

{\name} is an LLM-augmented visual analytics system for interdisciplinary collaborator discovery. Starting from a researcher's initial collaboration goal, the system translates it into structured tasks to facilitate broad exploration and cross-domain comparison. It then retrieves relevant literature from OpenAlex to ground these tasks in existing research, revealing how the goal connects to different academic fields. Through coordinated interactive visualizations, users can navigate candidate pools, trace research trajectories, and comparatively assess task alignment and complementary expertise. To bridge semantic divides prior to formal engagement, the system visualizes terminology differences and overlaps, while supporting publication-grounded conversation rehearsals. These capabilities help researchers evaluate collaboration feasibility and prepare for targeted initial outreach.

We evaluate {\name} through two case studies, a user study with 12 researchers, and a component-level evaluation. The results suggest that {\name} frames interdisciplinary collaborator discovery as a structured, evidence-based exploratory process. The case studies illustrate its use in assembling complementary teams and iteratively refining collaboration ideas. Furthermore, the user study indicates that the system increases transparency in early-stage exploration, supports the contextual comparison of candidate researchers, and boosts users' confidence in cross-domain communication. Taken together, these findings suggest that {\name} can help researchers transition from abstract collaborative intent to grounded candidate selection and informed initial contact.

The primary contributions of this paper are summarized as follows:
\begin{itemize}
    \item We characterize the problem space of interdisciplinary collaborator discovery through a formative study, distilling these empirical challenges into actionable design requirements.
    \item We present {\name}, an LLM-augmented visual analytics system for translating goals, exploring researchers, and preparing targeted outreach across domains.
    \item We evaluate {\name} through two case studies, a user study, and a component-level evaluation, providing empirical evidence of its effectiveness in complementary team formation, cross-domain idea refinement, and communication preparation.
\end{itemize}

\section{Related Work}
\subsection{Scientific Collaborator Recommendation}
Scientific collaborator recommendation \cite{husain2019expert} has transitioned from structural link prediction to semantic team formation. Early methods primarily leveraged topological features in co-authorship networks \cite{hassan2025supervised, yu2014predicting}. Techniques such as random walks, heterogeneous meta-paths (e.g., PathPredict \cite{sun2011co}, ArnetMiner \cite{tang2008arnetminer}), and random forest classifiers \cite{pavlov2007finding} were frequently used to predict future collaborations based on historical network proximity. However, these structural approaches are susceptible to homophily bias \cite{bekes2025cultural, fabbri2020effect}. They tend to recommend candidates from a researcher's existing social circle, which can limit interdisciplinary discovery.
To incorporate research content, subsequent studies framed academic matchmaking as a multi-constraint optimization problem \cite{berktacs2021branch, buyukboyaci2019team}, aiming to balance complementary expertise with network communication costs. While these methods utilize semantic profiles \cite{huang2021measuring}, they typically rely on coarse-grained, static aggregations of a scholar's publication history, which may not capture the specific requirements of a new project.

Although these methods are useful for identifying relevant scholars, they remain less suited to early-stage interdisciplinary exploration, where users may start with abstract goals rather than clear query terms or well-defined expertise needs. {\name} targets this missing step by translating collaboration intent into actionable, domain-specific tasks and grounding candidate retrieval in task-relevant literature.

\subsection{Visual Analytics of Academic Data}
Visual analytics is widely used to explore complex academic data \cite{fung2016design, heimerl2015citerivers, hao2022thirty}, supporting tasks such as literature analysis, bibliometric evaluation, and trend discovery \cite{dong2019vistory, guo2022sd, li2019galex}. Existing tools are generally categorized into macro-level science mapping \cite{narechania2021vitality} and micro-level scholar profiling \cite{dang2019wordstream, shi20141, wu2015egoslider}. For example, CiteSpace \cite{chen2006citespace} and VOSviewer \cite{van2010software} visualize broad disciplinary trends and citation networks. However, they often lack the granularity required to evaluate individual collaborators.

At the micro-level, ScholarPlot \cite{majeti2020scholar} provides visual summaries of academic achievements. To support deeper egocentric analysis, tools like Influence Flowers \cite{shin2019influence} map bi-directional citation influence, while GeneticFlow \cite{luo2023impact, xiao2023geneticflow} and GeneticPrism \cite{sun2025geneticprism} track research evolution alongside citation impact. Furthermore, interactive systems such as VISPubComPAS \cite{wang2019vispubcompas} and ACSeeker \cite{wang2021seek} enable scholarly exploration through user-guided faceted filtering. Recently, VizCV \cite{lazarik2025vizcv} incorporated AI-assisted analysis to model career trajectories based on topic evolution, impact metrics, and collaboration dynamics.

These systems offer valuable but mainly retrospective views of scholarly activity, focusing on what fields or scholars have done rather than whether a candidate fits a new collaboration goal. {\name} addresses this need through task-oriented views that connect user-specific tasks with temporal publication evidence and candidate comparison.

\subsection{Visual Analytics for Knowledge Alignment}
Visual analytics has been extensively employed to bridge semantic gaps and align domain-specific terminology across disciplines \cite{newman2021textessence, zhang2025anchortextvis}. Early text-based tools focused on revealing lexical differences across document facets or tracking temporal word frequency trends \cite{collins2009parallel, lee2010sparkclouds}. To support deeper semantic analysis, ConceptVector \cite{park2017conceptvector} leverages word embeddings to interactively construct concepts, while some other systems \cite{schatzle2017histobankvis, wu2026listeners, li2025conceptviz} map syntactic changes, model behaviors, or large language model (LLM) representations to human-understandable visual spaces. Beyond navigating terminological disparities, establishing cross-domain understanding requires broader visual sensemaking \cite{tergan2005use, sacha2014knowledge, weng2025insightlens}. SenseMap \cite{nguyen2016sensemap} and SensePath \cite{nguyen2015sensepath} capture this cognitive process by recording analytic provenance to help users curate fragmented information. Ultimately, this individual sensemaking must transition into collaborative grounding, where visualizations act as shared ``boundary objects'' \cite{luna2019modeling}. Foundational research demonstrates that shared visual interfaces and interaction minutiae are critical for establishing situational awareness, supporting consensus building, and reducing communication costs in cooperative environments \cite{heer2008design, gergle2004language, homaeian2021joint}.
While such work supports semantic analysis, sensemaking, and collaborative grounding, it largely focuses on knowledge artifacts or on interactions after collaboration begins. Before initial contact, researchers must interpret unfamiliar terminology and prepare for communication. {\name} targets this pre-collaboration stage with domain terminology comparison and publication-grounded rehearsal.

\section{Formative Study}
We conducted a formative study to investigate the practical challenges of initiating interdisciplinary collaboration and to inform the design of DeepConnect. Our goal was to understand how researchers articulate their requirements, search across domains, evaluate potential collaborators, and prepare for initial communication. Through analyzing archival collaboration requests and interviewing domain experts, we translated these observed barriers into concrete design requirements.

\subsection{Methodology and Participants}
Our study followed a two-phase qualitative approach, combining archival data analysis with expert interviews to systematically investigate this problem space.

First, we conducted a \textbf{thematic analysis} on an anonymized corpus of interdisciplinary collaboration proposals collected from a university-wide initiative. This corpus included requests from principal investigators across various STEM disciplines, including biological sciences, chemistry, physics, and food science, who were seeking external technical expertise. Analyzing these archival documents revealed how researchers naturally articulate their interdisciplinary needs.

Second, we conducted \textbf{semi-structured interviews} with six domain experts recruited through purposive sampling from our university. They included principal investigators and senior postdoctoral researchers from diverse STEM backgrounds, all with interdisciplinary collaboration experience or interest. Each interview lasted approximately 45--60 minutes and was conducted in person or online. The interviews examined collaborator search strategies, evaluation difficulties, and communication barriers. These expert interviews were covered by our university's Departmental Ethics Review Committee approval (no. 001325), and all interviewees provided informed consent. Because the interviews involved potentially sensitive early-stage collaboration ideas, we documented them through detailed written notes rather than audio recordings. Two authors analyzed the proposals and interview notes using open coding and synthesized key findings.

\subsection{Current Practice and Pain Points}
Synthesizing the archival data and expert interviews revealed a clear mismatch between the highly specific requirements of interdisciplinary collaboration and the limited support offered by existing exploratory workflows. We identified recurring bottlenecks spanning the entire collaboration process, from defining initial goals to preparing for contact.

\textbf{Coupling of Domain Context and Methodological Needs.} Our analysis indicated that interdisciplinary collaboration goals often intertwine domain context with implicit methodological requirements. Researchers rarely articulate their needs using standardized technical terms recognizable across disciplines. Instead, they rely on detailed, domain-specific narratives grounded in their immediate scientific problems. For example, a biologist seeking a spatial segmentation method might describe the morphological complexity of cellular structures, while a materials scientist requiring predictive modeling might emphasize chemical synthesis conditions. Experts noted that this domain-specific framing obscures the underlying computational tasks from outsiders. Consequently, potential collaborators often fail to recognize the relevance of their methods, even when they possess the exact expertise required.

\textbf{Bottlenecks in Current Discovery Channels.} When actively seeking potential collaborators, experts reported substantial friction across their primary discovery channels.

\underline{\textit{Limited Support from In-Person Academic Networking.}} Although academic events serve as a common starting point for interdisciplinary discovery, experts noted that such interactions are inherently unscalable and subject to high uncertainty. Informal conversations facilitate the exchange of broad interests but frequently fail to expose the task-level alignment necessary to determine project viability. Relying on physical networking restricts the search space to localized, serendipitous encounters, lacking the systematic reach required to comprehensively identify suitable collaborators for specific objectives.

\underline{\textit{Friction in Academic Search Platforms.}}
Traditional academic search engines impose a substantial cognitive burden. Experts described a persistent terminology mismatch: querying with familiar domain-specific terms predominantly returns immediate peers, whereas attempting to adopt unfamiliar terminology often yields an overwhelming number of irrelevant results. Furthermore, because these platforms are inherently document-centric, evaluating a candidate necessitates manual review of extensive publication records. This tedious process hinders the ability to form a holistic view of an individual's expertise, track its evolution over time, or compare multiple candidates against a unified collaboration goal.

\underline{\textit{The Exploratory Limits of Text-Based LLMs.}} Several experts reported utilizing LLMs to circumvent the limitations of traditional search. While these systems can mitigate terminology mismatches, text-based conversational interfaces exhibit clear limitations for exploratory discovery. Their linear responses are difficult to inspect systematically and fail to provide a macro-level overview of the researcher landscape, complicating multidimensional comparisons. Experts emphasized that discovering interdisciplinary partners is a high-stakes process demanding publication-grounded evidence and interactive exploration, which purely text-based interfaces cannot adequately fulfill.

\textbf{The Risks of Cross-Disciplinary Miscommunication.} While identifying a suitable collaborator is a critical step, it does not resolve the inherent difficulties of subsequent communication. Experts emphasized that differing academic paradigms frequently cause conceptual misalignment. Specifically, domain-specific terminologies and habitual expressions act as significant barriers to mutual understanding, resulting in highly inefficient dialogue. Furthermore, evaluating the feasibility of a proposed idea during initial discussions is inherently difficult. Due to a lack of cross-domain familiarity, researchers often underestimate the methodological complexity of a task within the target discipline, or conversely, overlook potentially fruitful collaborative opportunities. Therefore, researchers highlighted a critical need to validate ideas and establish shared context prior to initial contact.

\subsection{Design Requirements}
To address these challenges, we distilled four design requirements (R1--R4) for supporting interdisciplinary collaboration initiation.

\textbf{R1: Translate Collaboration Goals into Methodological Tasks.} To isolate technical needs from domain-specific contexts, the system must translate broad collaboration goals into actionable tasks. It should let researchers express intentions naturally and map them to methodological requirements recognizable across disciplines.

\textbf{R2: Enable Comparative Evaluation of Researcher Fit.} To alleviate the cognitive burden of manual candidate evaluation, the system must provide multidimensional visual evidence of a researcher's expertise. It should summarize an individual's academic trajectory, explicitly assess their relevance to the extracted tasks, and support visual comparisons of complementary strengths among candidates.

\textbf{R3: Reveal Terminology Gaps and Overlaps Across Domains.} To mitigate friction caused by unfamiliar vocabularies, the system should visualize terminological differences and alignments between disciplines. By highlighting shared concepts and clarifying domain-specific expressions, the system can help users build shared context before communication.

\textbf{R4: Enable Expertise-Grounded Conversation Rehearsal.} To minimize unproductive discussions, the system must offer a safe, interactive environment to prepare for initial contact. It should enable users to systematically test the methodological feasibility of their ideas against a candidate's domain expertise, allowing researchers to iteratively refine their communication strategies.

\section{System Overview}

\begin{figure*}[t]
 \centering
 \includegraphics[width=\linewidth]{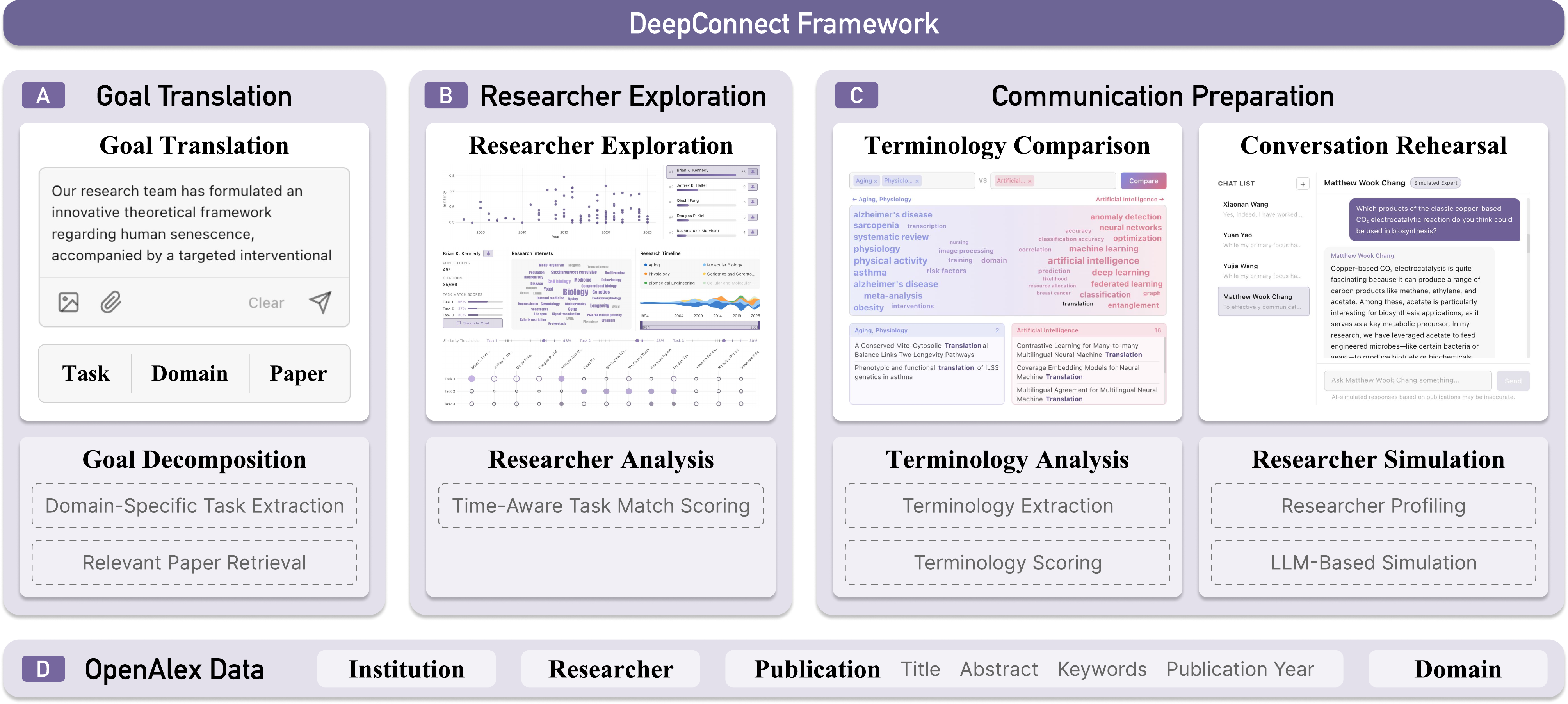}
 \caption{The visual analytics framework of DeepConnect. The system comprises four coordinated modules: (A) Goal Translation decomposes collaborative objectives into domain-specific Tasks and retrieves Relevant Papers; (B) Researcher Exploration supports the macroscopic overview, detailed profiling, and comparative assessment of candidate scholars; (C) Communication Preparation reveals terminological overlaps and enables publication-grounded Conversation Rehearsals to bridge cognitive gaps; and (D) the foundational OpenAlex Data that grounds the aforementioned three modules by connecting institutions, researchers, publications, and domains.}
 \label{fig:framework}
\end{figure*}

{\name} is a visual analytics system designed to facilitate interdisciplinary collaborator discovery using scholarly data from OpenAlex (Fig. \ref{fig:framework}D).
The workflow is organized around the requirements identified in the formative study. It begins with \textbf{Goal Translation} (Fig. \ref{fig:framework}A), where we use an LLM to translate initial goals into actionable domain-specific tasks and retrieve cross-domain literature (\textbf{R1}). These tasks guide \textbf{Researcher Exploration} (Fig. \ref{fig:framework}B), which combines temporal scatterplots, detailed profiles, and comparative matrices to examine evolving expertise and task fit (\textbf{R2}). \textbf{Communication Preparation} (Fig. \ref{fig:framework}C) visualizes terminology gaps and overlaps to establish common ground (\textbf{R3}) and provides publication-grounded conversation rehearsal for feasibility exploration and communication planning (\textbf{R4}).

\subsection{Data Collection and Processing}
We use OpenAlex\footnote{\url{https://openalex.org/}}, an open-access catalog of scholarly entities, as the primary data source for our system. OpenAlex represents scholarly data through interconnected entities, including \textit{Institution}, \textit{Author} (researcher), \textit{Work} (publication), and \textit{Domain}. Within this structure, authors are linked to their affiliated institutions, while works are linked to their contributing authors and associated domains. Each Work provides publication-level metadata, including its title, abstract, keywords, and publication year. These data provide the basis for semantic analysis, terminology comparison, and temporal trajectory tracking.

To ground our system in a practical application scenario, we constructed a representative local-university subset containing affiliated researchers and their associated publications. Facilitating collaborations within a single institution lowers administrative and spatial barriers, making the identified interdisciplinary opportunities more actionable and realistic for researchers. We use this curated dataset throughout our case studies and user study to evaluate the system's practical utility.

\subsection{Goal Decomposition}
\label{sec:goal-decomposition}
We design the Goal Decomposition module to transform open-ended collaboration goals into actionable domain-specific tasks and retrieve relevant papers to ground subsequent exploration (\textbf{R1}).
To accommodate different stages of research ideation, DeepConnect accepts multimodal input, including textual descriptions, conceptual images, and project proposal documents.

\textbf{Domain-Specific Task Extraction.}
We use GPT-5.1 to derive up to five semantically diverse tasks that cover the collaboration goal and support systematic cross-domain search.

\textbf{Relevant Paper Retrieval.}
To guide candidate search, we ground the extracted tasks in relevant literature.
We use the all-MiniLM-L6-v2 embedding model to encode both the extracted tasks and the OpenAlex paper metadata (paper title, abstract, and keywords) into a shared semantic space \cite{sun2024framework, feng2025dont, sun2026implicit, sun2025swallow}. The system then computes the cosine similarity between the task embeddings and paper embeddings to rank the literature.

\subsection{Researcher Analysis}
\label{sec:researcher-analysis}
We identify candidate researchers by extracting authors from the literature retrieved for each task $t$.
Sourcing candidates directly from these relevant publications ensures that each scholar is associated with verifiable scholarly metadata, including specific papers, publication years, and research domains, which provides a reliable foundation for subsequent analysis (\textbf{R2}).

We use \textbf{Time-Aware Task Match Scoring} to estimate the alignment between a candidate researcher $r$ and a task $t$ based on both semantic relevance and publication recency. As research interests evolve over time, recent publications indicate current focus, while earlier works reflect foundational expertise.
For each paper $p \in P_r$, we calculate a relevance score $Sim(t,p)$ using the cosine similarity between the task and paper embeddings. This score is then adjusted by a temporal weight $w(p)$ based on the publication year $Y_p$:
\begin{equation}
w(p) = \alpha + (1-\alpha)\exp\left(-\lambda (Y_{\max} - Y_p)\right)
\end{equation}
Here, $Y_{\max}$ represents the latest publication year in the dataset, $\lambda$ determines the decay rate, and $\alpha \in [0,1]$ establishes a lower bound for the weight of older publications (set to $0.5$ in our implementation).

Let $Top_x(P_r,t)$ represent the set of $k$ top-ranked papers in $P_r$ based on the weighted relevance score $w(p) \cdot Sim(t,p)$, where $k = \min(x, |P_r|)$ and $x$ defaults to $5$. The final match score is defined as:
\begin{equation}
\mathit{Score}(r,t) = \frac{1}{k} \sum_{p \in \mathit{Top}_x(P_r,t)} w(p) \cdot \mathit{Sim}(t,p)
\end{equation}

We aggregate only the top-matched papers to avoid diluting task-specific relevance with unrelated historical topics, especially for senior researchers with diverse publication records.
This averaging approach supports robust ranking, filtering, and evaluation of candidates by capturing both sustained and current expertise.

\subsection{Terminology Analysis}
\label{sec:terminology-analysis}
To prepare users for target-domain concepts, we analyze domain terminology to reveal unfamiliar terms and shared vocabulary across fields (\textbf{R3}).
This process consists of three steps.

\textbf{Domain-Specific Terminology Extraction.} We first construct a text corpus for domain terminology comparison. For each domain, the system retrieves the top 1000 most relevant papers using the embedding model, a threshold chosen to balance literature coverage with real-time performance. Grounded in this academic context, we deploy a lightweight LLM (GPT-4.1-mini) to extract conceptual terminology from the titles and abstracts, ensuring high-speed processing across thousands of documents.
The extraction is designed to retain representative technical terms that reflect the conceptual focus of the domain.
To keep the resulting terminology set analytically useful, the extraction excludes variable names, proper nouns, and generic academic verbs.

\textbf{Terminology Scoring.} After extraction, the system estimates the importance of each term with a domain-level TF-IDF formulation \cite{sparck1972statistical,salton1988term}. For a term $w$ in the domain $D_A$, the term frequency $TF(w, D_A)$ measures how frequently $w$ is observed in the retrieved literature of $D_A$.
The inverse document frequency is computed over the global set of retrieved papers $P_{\mathrm{all}}$ to downweight terms that broadly appear across many papers.
The resulting importance score $S(w, D_A)$ is defined as
\begin{equation}
S(w, D_A) = \mathit{TF}(w, D_A) \cdot \log \left(
\frac{|P_{\mathrm{all}}|}
{\left| \left\{ p \in P_{\mathrm{all}} : w \in p \right\} \right|}
\right)
\end{equation}
Here, $\left| \left\{ p \in P_{\mathrm{all}} : w \in p \right\} \right|$ denotes the number of papers in which the term $w$ appears.

To quantify terminology overlap across domains, the system further computes a relative frequency ratio for each extracted term. Given the user domain $D_B$ and a comparison domain $D_A$, let $Freq(w, D)$ denote the number of retrieved papers in domain $D$ that contain term $w$. The relative ratio $R(w)$ is defined as
\begin{equation}
R(w) =
\frac{\mathit{Freq}(w, D_A)}
{\mathit{Freq}(w, D_A) + \mathit{Freq}(w, D_B)}
\end{equation}

This metric normalizes the term distribution within the range $[0, 1]$. A value close to $1$ indicates that the term is used primarily in $D_A$, while a value close to $0$ indicates that it is used primarily in $D_B$. A value near $0.5$ indicates shared terminology that appears with similar frequency in both domains. This metric provides the quantitative basis for the horizontal placement of terms in the Terminology Comparison view.

\subsection{Publication-Grounded Researcher Simulation}
To support feasibility exploration before initial contact, we provide an interactive, publication-grounded simulation for rehearsing questions and anticipating misunderstandings (\textbf{R4}).

\textbf{Researcher Profiling.}
Upon selecting a candidate researcher $r$, the system aggregates their key metadata to construct a bounded grounding context $\mathcal{C}$. This context incorporates the researcher's associated domains ($\mathit{Domains}_r$) and the abstracts of their complete publication record ($P_r$) to capture their academic trajectory and expertise.

\begin{equation}
\mathcal{C} = \{ \mathit{Name}_r, \mathit{Domains}_r, \mathit{Abstracts}(P_r) \}
\end{equation}

\textbf{LLM-Based Simulation.} By conditioning an LLM (GPT-5.1) on this localized context, the generation probability of any response $R$ given a user query $Q$ is explicitly modeled as $P(R \mid Q, \mathcal{C})$.
To mitigate generative hallucinations, we impose strict semantic boundaries on the simulation \cite{xu2024hallucination}.
The model is configured to ground its outputs exclusively in the provided publication records, reflect the candidate researcher's established academic perspective, and acknowledge topics falling outside their documented expertise \cite{zhang2025treeqa}. This evidence-based architecture ensures a reliable preparation environment for communication across domains.

\section{Visual Analytics System}
DeepConnect guides users from open-ended collaboration goals to actionable partnerships through an interactive visual analytics workflow. The exploratory process begins by translating user inputs into domain-specific tasks to establish a foundational literature base. With a structured task representation, users systematically evaluate and compare candidate researchers through coordinated visual evidence. Once promising candidates are identified, the system facilitates communication preparation by visualizing terminology gaps and supporting publication-grounded conversation rehearsal before initial contact.

\subsection{Collaboration Goal Translation}
The \textbf{Goal Translation} (\autoref{fig:system}A) view is designed to structure users' initial collaboration goals and connect them to verifiable literature evidence.
Users first articulate their objectives in the \textit{Research Goal} panel. The system processes this input and generates tabs for each extracted domain-specific task. These tabs serve as global contextual filters. Selecting a \textit{Task} updates subsequent visual components to reflect the active analytical focus.

To provide a macroscopic perspective on the interdisciplinary nature of the active task, the \textit{Domain Distribution} panel uses a horizontal bar chart to summarize the domain distribution of the top 100 retrieved publications. Interactive filtering is naturally supported: clicking a specific domain bar updates the adjacent \textit{Relevant Papers} panel to display only literature from the selected field.

The \textit{Relevant Papers} panel displays the retrieved seed publications, detailing their task relevance scores, publication years, and domain labels. Users can expand individual items to read full abstracts or navigate to the external Google Scholar website. When a candidate is selected later, this list updates to show only their task-matched publications.

\subsection{Contextual Researcher Exploration}
The \textbf{Researcher Exploration} (\autoref{fig:system}B) view coordinates three vertically arranged panels to support both macroscopic discovery and individual-level profiling.

The \textit{Researcher Overview} (\autoref{fig:system}B1) panel supports initial candidate assessment. It features a scatterplot mapping publication years to the x-axis and embedding similarity scores to the y-axis, alongside a ranked list of candidate researchers sortable by publication volume or task match scores (Section 4.3). This panel tightly couples cross-view interactions. Within the scatterplot, brushing a specific region to isolate papers dynamically recalculates the candidate rankings, and clicking a data point links directly to the underlying literature in the Relevant Papers panel. Conversely, within the ranked list, hovering over a candidate researcher highlights their publications in the scatterplot to reveal temporal work patterns, and selecting a candidate updates the subsequent Relevant Papers panels for deeper inspection.

The \textit{Researcher Details} (\autoref{fig:system}B2) panel offers a microscopic profile of the selected candidate researcher. Alongside basic academic metadata and aggregate task match scores, a \textit{Research Interests} word cloud maps keyword frequency to text size to reveal the individual's overarching focus. Adjacent to this, a \textit{Research Timeline} streamgraph encodes publication volume into stream width to visualize their evolutionary trajectory across disciplines. To guide analytical focus, keywords and streams highly relevant to the active task are highlighted, whereas irrelevant ones are in gray text or semi-transparency. If a candidate appears suitable, users can initiate a simulated dialogue in the Conversation Rehearsal view.

The \textit{Researcher Comparison} (\autoref{fig:system}B3) panel features an interactive dot matrix to help users balance expertise across disciplines. Rows represent the extracted tasks, while columns display candidate researchers pinned during the exploration phase. Dot size and opacity visually encode a candidate researcher's match score for the corresponding task. By interactively adjusting independent score threshold sliders, dots meeting the criteria remain solid, whereas those falling below are rendered as hollow circles.
This visual encoding makes task-level expertise gaps visible and supports users in comparing whether selected candidates collectively cover the task requirements.

\subsection{Terminology Comparison}
To bridge the semantic gap, the \textbf{Terminology Comparison} (\autoref{fig:system}C) view maps terminology gaps and overlaps across domains. Based on the relative frequency ratio calculated in Section 4.4, a word cloud distributes the extracted terms along a horizontal axis. Terms positioned near the lateral edges represent domain-specific terminology that users should approach with caution or learn in advance to prevent misunderstandings. Conversely, centrally located terms indicate shared terminology that helps establish common ground. To reinforce this visual mapping, a color gradient transitions from blue on the left to red on the right, with shared terms blending into purple to symbolize the conceptual convergence of the compared domains.

To support contextual traceability, the system links these terms directly to the underlying literature. When a specific term is selected, the bottom panel displays sentences from the literature containing that term. This enables users to examine how unfamiliar domain terminology is practically applied in the target discipline's literature.

\subsection{Conversation Rehearsal}
The \textbf{Conversation Rehearsal} (\autoref{fig:system}D) view adopts a familiar messaging layout, featuring a \textit{Chat List} on the left for managing multiple sessions and a main dialogue window on the right. To establish analytical trust and mitigate over-reliance on the LLM, the system marks the candidate researcher as a simulated expert and explicitly disclaims that all responses are derived from their past publications (Section 4.5). Reinforcing this factual grounding, the session initiates with a self-introduction generated directly from these academic records.
This environment supports communication rehearsal, conceptual clarification, and feasibility exploration before contact.

\section{Case Study}
To evaluate the effectiveness and utility of \textit{DeepConnect} in real-world exploratory scenarios, we present two case studies. These cases involve $E1$ and $E2$, corresponding to user-study participants $P11$ and $P12$, respectively. Both used the system to address their actual interdisciplinary challenges. The first case focuses on task-driven team assembly for an AI-driven biohybrid system, while the second explores semantic alignment and iterative refinement in a sensor-vision fusion project. Together, they show how the system connects broad planning with domain-specific grounding.

\subsection{Task-Driven Complementary Team Assembly}
$E1$ is an environmental engineering researcher specializing in electrocatalytic $\mathrm{CO}_2$ reduction. To overcome the efficiency bottlenecks of single chemical conversion, $E1$ aimed to develop a cascaded system integrating upstream electrocatalysis with downstream biomanufacturing, regulated autonomously by machine learning algorithms. This objective required an interdisciplinary team spanning materials science, synthetic biology, and artificial intelligence. $E1$ used \textit{DeepConnect} to break down the project goal and identify researchers whose expertise could jointly support these needs.

\textbf{Goal Translation and Task Grounding.} $E1$ first input this high-level goal into the Goal Translation view (\autoref{fig:system}A). The system decomposed the request into three domain-specific tasks: (1) electrocatalytic reaction engineering for optimized $\mathrm{CO}_2$ reduction processes, (2) synthetic biology approaches for scalable biomanufacturing, and (3) autonomous control systems using machine learning for real-time reactor optimization. $E1$ used these tasks to systematically explore potential collaborators.

\begin{figure}[t]
 \centering
 \includegraphics[width=\linewidth]{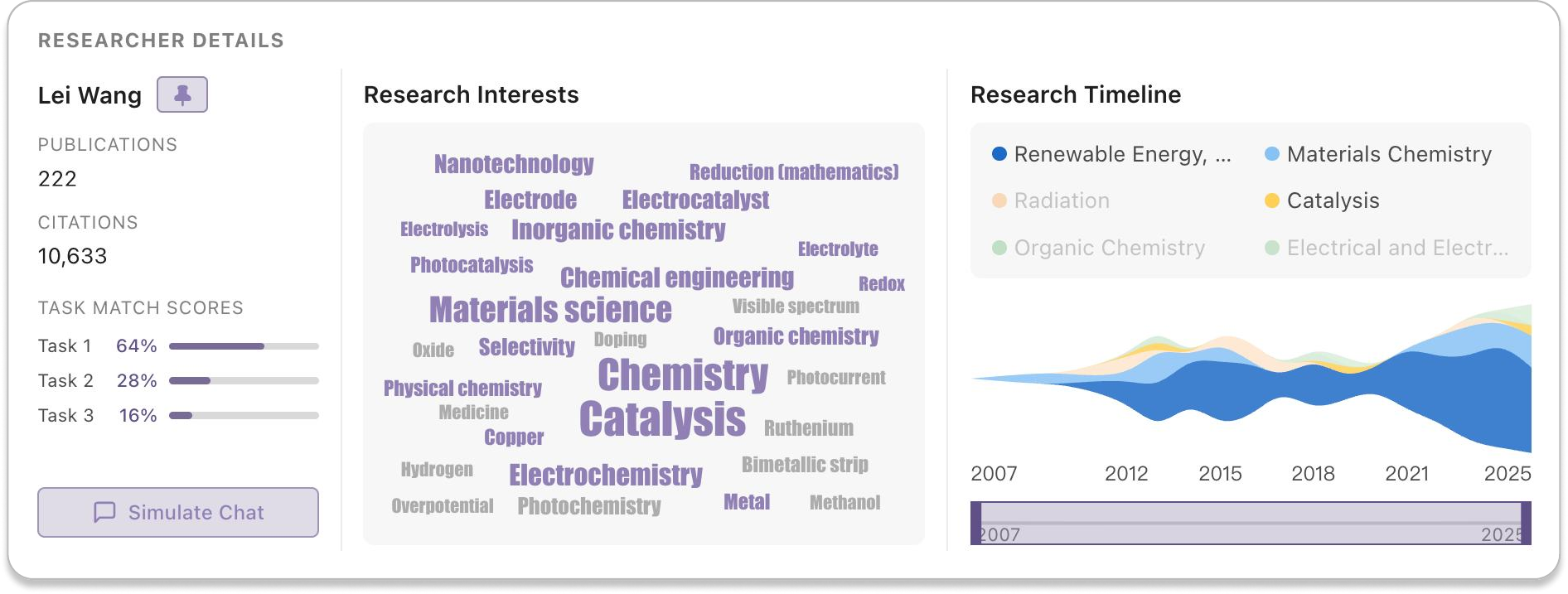}
 \caption{\textbf{Case 1:} The Researcher Details panel illustrates Prof. Lei Wang's expertise for the electrocatalytic reaction engineering task. The Research Interests and Research Timeline confirm sustained, task-relevant materials science output for the task context.}
 \label{fig:case-1-1}
\end{figure}

\textbf{Contextual Profiling and Collaborator Identification.} Focusing on the first task, $E1$ found that the retrieved literature in the Domain Distribution was concentrated in renewable energy and sustainability. Although already familiar with reactor design, $E1$ needed a materials scientist with strong expertise in catalyst development. In the ranked list of the Researcher Overview view, $E1$ identified Prof. Lei Wang, who exhibited the highest matching score and publication count. To examine this candidate more closely, $E1$ turned to the Researcher Details view (\autoref{fig:case-1-1}). The Research Interests word cloud emphasized materials-related keywords such as \textit{Electrochemistry}, \textit{Electrocatalyst}, and \textit{Carbon monoxide}. The Research Timeline streamgraph further showed sustained recent output in renewable energy. Taken together, these visual cues indicated a strong and active research focus aligned with the task, leading $E1$ to pin Prof. Lei Wang.

\textbf{Visual Semantic Alignment and Conversation Rehearsal.} Proceeding to the second task, $E1$ noted from the Domain Distribution a focus on molecular biology and biotechnology. Since this field fell outside their expertise, $E1$ compared candidates using the scatterplot in the Researcher Overview (\autoref{fig:system}B1). Prof. Matthew Wook Chang emerged as a strong candidate due to his high match score and recent publication volume, with the Research Timeline (\autoref{fig:system}B2) verifying his sustained engagement in molecular biology. To prepare cross-disciplinary communication, $E1$ inspected the Terminology Comparison (\autoref{fig:system}C) for conceptual alignments and gaps. Unfamiliar biological concepts (e.g., \textit{metabolic flux}, \textit{enzymatic hydrolysis}) indicated areas requiring preliminary study. Meanwhile, domain-specific electrochemistry terms (e.g., \textit{Faradaic efficiency}) prompted $E1$ to consider how to explain them in language accessible to a biologist. Shared concepts like \textit{Syngas production} established a practical conversational entry point. Guided by this semantic preparation, $E1$ used the Conversation Rehearsal (\autoref{fig:system}D) to discuss biological carbon sources and electrochemical metrics in accessible terms.

\begin{figure}[t]
 \centering
 \includegraphics[width=\linewidth]{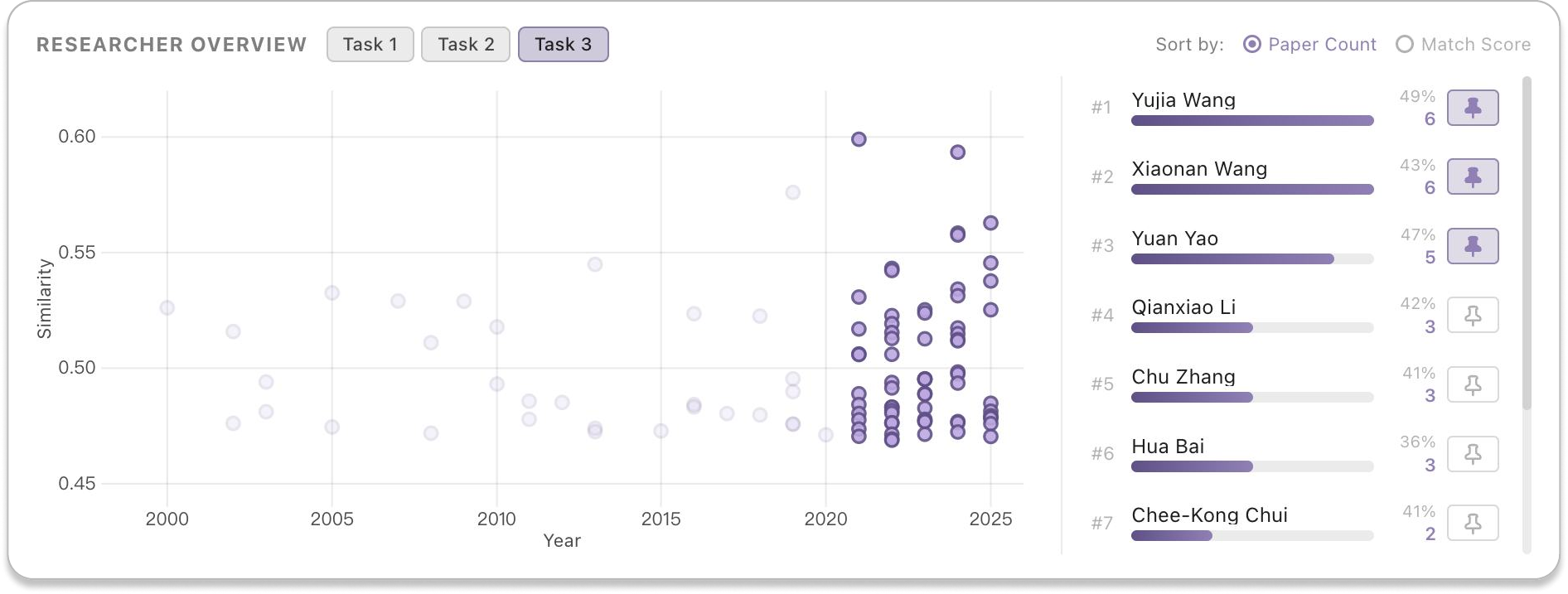}
 \caption{\textbf{Case 1:} Temporal filtering within the Researcher Overview for the artificial intelligence Task. Brushing the scatterplot to isolate recent publications dynamically updates candidate rankings, enabling the user to identify active researchers such as Prof. Xiaonan Wang.}
 \label{fig:case-1-2}
\end{figure}

\textbf{Interactive Visual Filtering and Qualitative Assessment.} For the third task, the target literature centered on control systems and artificial intelligence. Observing that the initially top-ranked candidate primarily held older publications, $E1$ brushed the scatterplot to filter for recent academic works (\autoref{fig:case-1-2}). This dynamic update revealed three researchers with comparable relevance scores and publication volumes. $E1$ pinned these candidates to conduct comparative conversation rehearsals. During discussions on parameter optimization, the publication-grounded simulated response for Prof. Xiaonan Wang offered the most relevant insights, highlighting work on machine learning for complex chemical reactions such as hydrothermal processes. This indicated robust dual expertise in algorithms and chemical engineering, prompting $E1$ to select her for the AI-control task.

\textbf{Visual Comparison and Team Assembly.} Finally, $E1$ examined the Researcher Comparison (\autoref{fig:system}B3). After adjusting the similarity threshold sliders, the dot matrix confirmed that the three selected candidates (Prof. Lei Wang, Prof. Chang, and Prof. Xiaonan Wang) provided complementary and comprehensive coverage across the three tasks. Moreover, the matrix revealed that Prof. Xiaonan Wang exhibited higher relevance for the first two tasks compared to her peers. This visual pattern indicated a strong cross-disciplinary background, which could effectively lower communication barriers during the trilateral collaboration. Supported by this visual and conversational evidence, $E1$ finalized the team assembly and proceeded to draft formal invitations.

\subsection{Bridging Cognitive Gaps for Sensor-Vision Fusion}
$E2$, a researcher in mechanical and biomedical engineering, developed a soft robotic exoskeleton glove equipped with high-density strain sensors and an electromyography (sEMG) module. A key bottleneck in this system was that physical sensor-based grasp intent recognition remained highly susceptible to environmental noise and user variance. To address this issue through artificial intelligence, $E2$ utilized \textit{DeepConnect} to discover cross-disciplinary expertise.

\textbf{Initial Exploration and Conversation Rehearsal.} $E2$ inputted her collaborative objective into the Goal Translation: ``I developed a soft robotic exoskeleton. I want to apply advanced deep learning algorithms to process my sEMG and soft strain sensor data to improve the accuracy of grasp intent recognition.'' Following the system's task extraction, $E2$ reviewed the recommended candidates and selected Prof. Angela Yao due to her extensive publication record in deep learning and complex sequence processing. Seeking algorithmic guidance, $E2$ initiated a Conversation Rehearsal to determine which models could enhance intent classification for noisy, user-dependent data. The simulated expert provided actionable insights, suggesting models capable of temporal aggregation (e.g., C2F-TCN) to capture long-range dependencies, alongside feature disentanglement techniques to isolate motion intent from user-specific noise.

\begin{figure}[t]
 \centering
 \includegraphics[width=\linewidth]{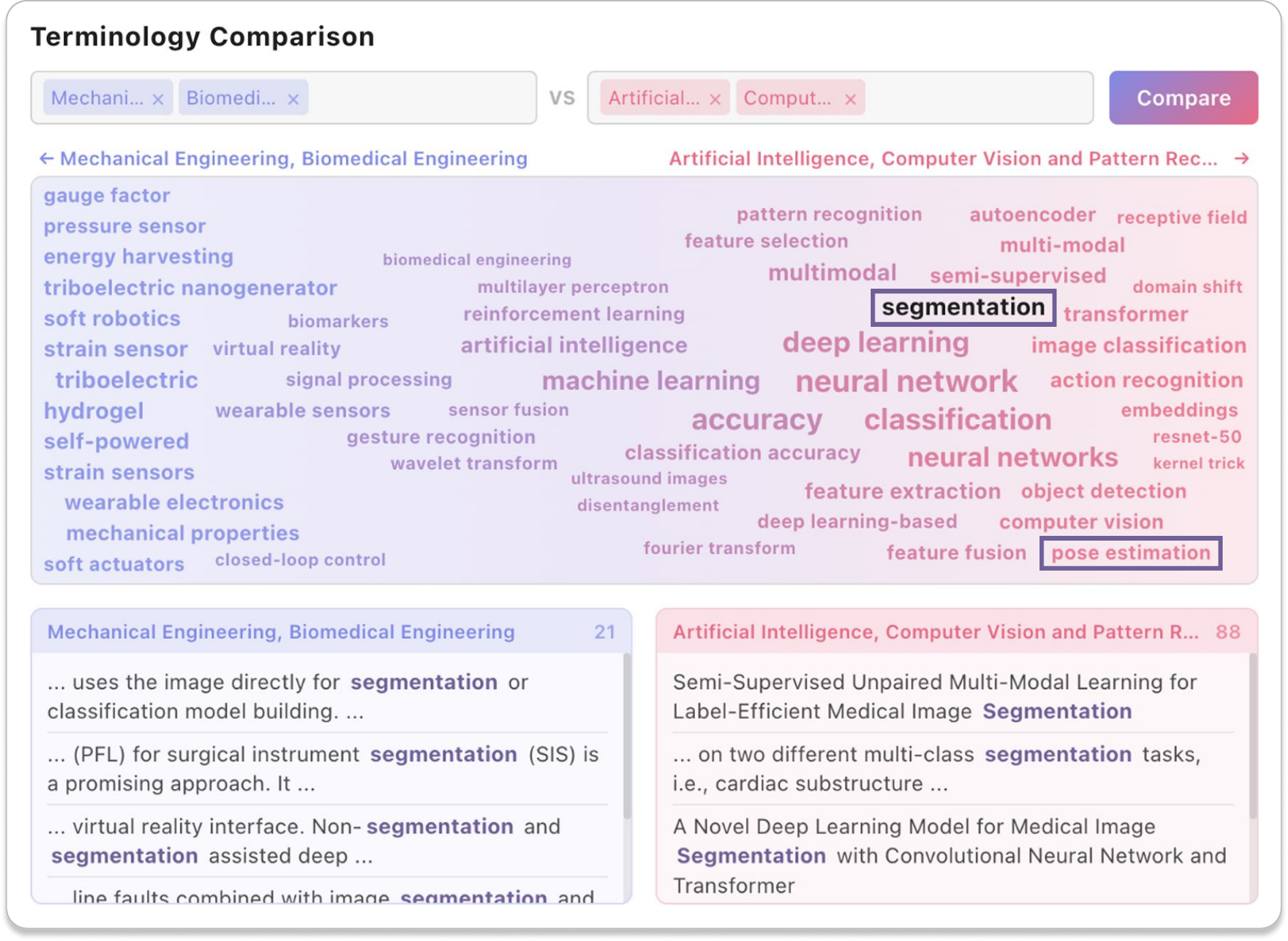}
 \caption{\textbf{Case 2:} Semantic alignment within the Terminology Comparison for grasp intent recognition. Highlighting terminological gaps across fields related to mechanical engineering and artificial intelligence reveals some complementary concepts such as ``segmentation'' and ``pose estimation'', prompting a novel sensor-vision fusion approach.}
 \label{fig:case-2}
\end{figure}

\textbf{Visual Sensemaking and Cognitive Shift.} Acknowledging feature extraction and disentanglement as promising technical directions, $E2$ sought to bridge her knowledge gap regarding these computational concepts. To understand how physical sensor data aligned with these algorithms, $E2$ examined the Terminology Comparison (\autoref{fig:case-2}). While her domain on the left was anchored by familiar terms like \textit{strain sensors} and \textit{wearable sensors}, $E2$ noticed that the expert's domain on the right predominantly featured computer vision terminology such as \textit{segmentation} and \textit{pose estimation}. This observation prompted a critical realization. She deduced that temporal disentanglement algorithms might underperform if applied exclusively to one-dimensional physical signals. Since grasping inherently involves spatial interaction with the environment, $E2$ recognized that visual information would provide spatial context missing from her current physical sensors.

\textbf{Iterative Refinement.} Building on this insight, $E2$ hypothesized that integrating a micro-camera to capture spatial priors could significantly simplify the noise processing challenge. To explore this architectural shift, she returned to the Conversation Rehearsal. $E2$ proposed a new framework that combined the expert's feature extraction methods with visual inputs, suggesting the use of lightweight \textit{segmentation} or \textit{pose estimation} models to capture macroscopic environmental features for subsequent \textit{feature fusion} with the microscopic strain data. The simulated response suggested that visual priors effectively compensate for the spatial blind spots of physical sensors. Through this iterative process, \textit{DeepConnect} facilitated a crucial pivot from a unimodal algorithmic query to a comprehensive vision-motor fusion strategy.

\section{User Study}
We conducted a user study to assess {\name}'s effectiveness and usability in interdisciplinary matchmaking. The study examined how researchers used the visual analytics workflow for cross-domain exploration and collected qualitative feedback on the system design.
The user study was covered by the same ethics approval (no. 001325).

\subsection{Participants and Apparatus}
We recruited 12 researchers ($P1-P12$) from a local university through email invitations and targeted outreach, including one professor, four postdoctoral researchers, five doctoral candidates, and two master's students. All were actively engaged in academic research, with research experience ranging from approximately 1 to 10 years. Their domains included computer and information sciences (AI, HCI, and clinical informatics), earth and environmental sciences (climate change, carbon cycling, hydrology, and environmental engineering), physical and life sciences (materials science, chemical biology, biomedical engineering, and bioenergy), and behavioral sciences (psychology). All had experience or strong interest in interdisciplinary research.

Each one-on-one study session lasted approximately 60 minutes. To accommodate individual preferences, sessions occurred either in person using a provided laptop or remotely via video conferencing with screen control capabilities. All participants provided informed consent covering participation, audio recording, and screen capture. Upon completion, each individual received a compensation of SGD~20.

\subsection{Study Procedure}
The evaluation comprised four structured phases.

\textbf{Introduction and Background Survey (10 minutes).} The session began with the project background and research motivation behind {\name}. Participants then completed a brief questionnaire on academic domains, research experience, and search strategies.

\textbf{System Tutorial (10 minutes).} An instructor guided participants through the {\name} interface using a predefined usage scenario. This tutorial explained the visual encodings, interactions, and data processing logic across the core modules. Participants then explored sample data to learn the workflow before the task.

\textbf{Open-Ended Exploration (20 minutes).} During the core task phase, participants formulated a realistic interdisciplinary idea aligned with their current interests. Using this objective, they used {\name} to discover, evaluate, and simulate communication with potential collaborators. Participants followed a think-aloud protocol, verbalizing their strategies and reasoning. The instructor observed silently and intervened only when clarification was needed.

\begin{figure}[t]
 \centering
 \includegraphics[width=\linewidth]{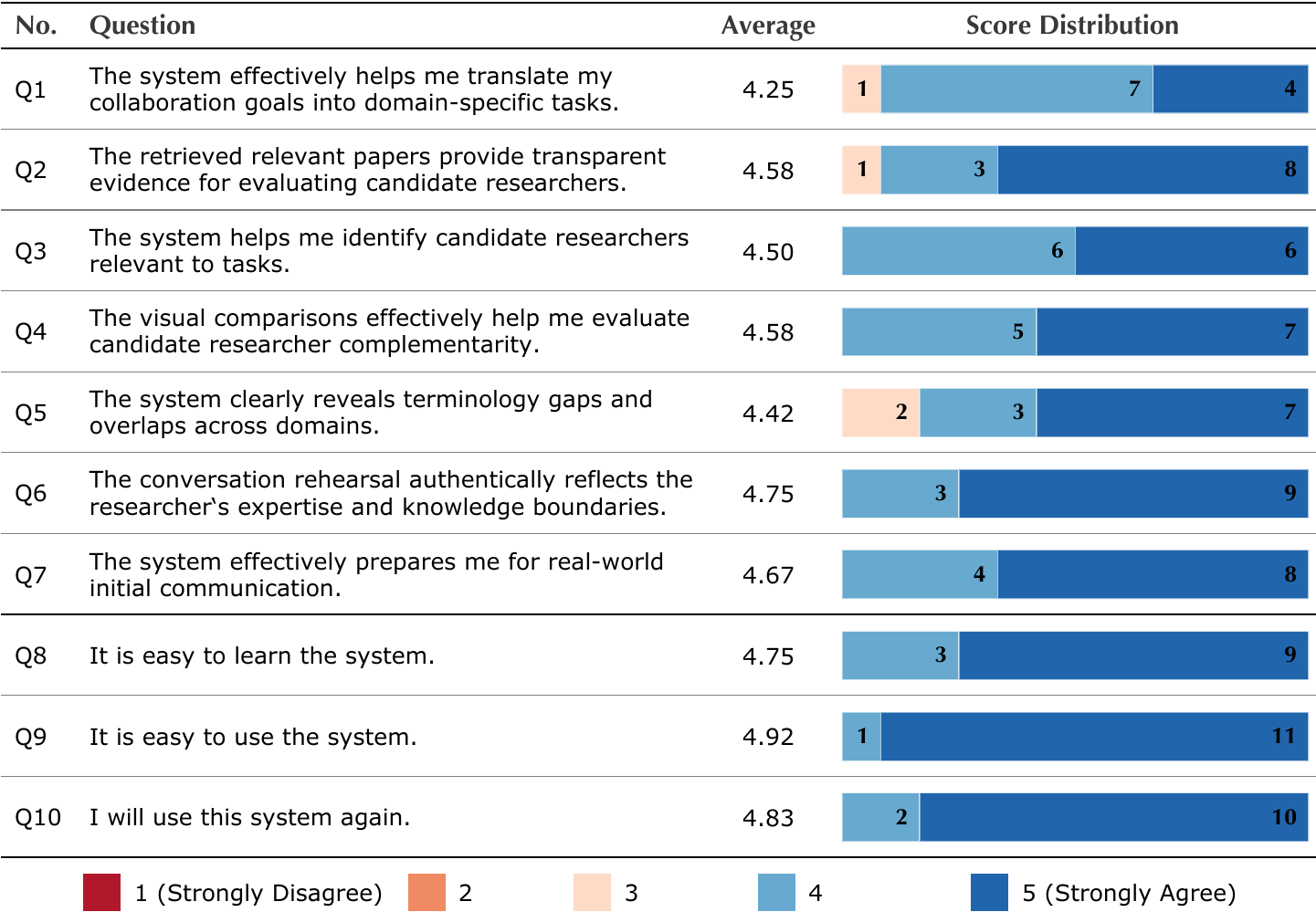}
 \caption{Quantitative user-study results from the 5-point Likert-scale questionnaire. The score distributions indicate consistently positive participant feedback on the system's effectiveness in supporting intent translation, expertise evaluation, cognitive alignment, and overall usability.}
 \label{fig:evaluation}
\end{figure}

\textbf{Post-Study Feedback and Interview (20 minutes).} Following the exploration, participants completed a 5-point Likert questionnaire (\autoref{fig:evaluation}) assessing the system's effectiveness and usability. Finally, a semi-structured interview gathered rating rationales and design input.

\subsection{Qualitative Feedback and Observations}
The ratings and semi-structured interviews provided insights into the system's utility and participants' analytical processes. We organize the feedback by workflow stage.

\subsubsection{Goal Translation and Evidence Transparency}
Participants highly rated the system's ability to \textbf{translate collaboration goals into domain-specific tasks} (Q1: 4.25/5). $P2$ noted that the generated tasks bridged familiar domain objectives with target discipline requirements. $P6$ also confirmed the accuracy of the extracted tasks, stating, ``The system actually reminded me of potential research directions that I had overlooked.'' However, $P12$ noted that some subtasks emphasized different aspects than expected. When the analytical focus of the extracted tasks required adjustments, $P7$ and $P12$ found the system highly flexible. They could easily guide the model to output more aligned task structures by refining their input descriptions.

Regarding the \textbf{exploration transparency} (Q2: 4.58/5), participants confirmed that the \textit{Relevant Papers} established a robust foundation for exploration. $P3$ emphasized that visualizing these relevant papers offered crucial interpretability by justifying the high rankings of specific researchers. Furthermore, $P7$ observed that displaying publication years alongside these papers enabled users to quickly distinguish a candidate's recent active focus from past endeavors.

\subsubsection{Researcher Evaluation}
Participants confirmed that the \textit{Researcher Exploration} effectively \textbf{identified researchers relevant to the decomposed tasks} (Q3: 4.50/5). An interesting pattern emerged regarding the profiles of recommended candidates. $P1$ observed that the system successfully surfaced boundary-spanning researchers with established dual expertise at the intersection of the targeted domains. For instance, $P6$, a medical researcher seeking an artificial intelligence collaborator, noticed that the system prioritized artificial intelligence scholars with medical crossover experience over purely theoretical computer scientists. $P6$ stated, ``This pragmatic recommendation is extremely helpful for actual research, as it lowers the communication barrier and facilitates much more aligned collaborations.'' $P3$ suggested impact-factor filters and author-position indicators for assessing academic influence.

Participants confirmed that the visual comparisons were highly effective for \textbf{evaluating candidate complementarity} (Q4: 4.58/5). $P1$ commended the \textit{Researcher Comparison}, emphasizing its substantial utility in constructing well-balanced collaborative groups. $P2$ contrasted this visual approach with traditional conversational interfaces, noting that standard text-based models demand repetitive prompting to extract and compare candidate profiles. The interactive matrix made complementary strengths immediately visible. $P10$ further suggested automatically recommending expert combinations for team assembly.

\subsubsection{Communication Preparation}
Participants confirmed that the \textit{Terminology Comparison} effectively supported cross-disciplinary preparation by \textbf{revealing conceptual differences across domains} (Q5: 4.42/5). $P6$ noted that the visual encoding exposed ontological variations, contrasting the broad definition of artificial intelligence in public health with its specific sub-fields in computer science. To enhance this feature, $P11$ suggested brief LLM-generated definitions for displayed terms.

Participants widely utilized the \textit{Conversation Rehearsal} to probe \textbf{candidates' knowledge boundaries} (Q6: 4.75/5). Users frequently employed a trust-building strategy by initially simulating chats with familiar researchers. After reviewing these publication-grounded responses, participants proceeded to interact with unfamiliar candidates. $P8$ appreciated that the simulated researchers explicitly acknowledged their research boundaries while delivering comprehensive answers within their domain. Highlighting the efficiency of this feature, $P12$ stated, ``Using simulated chats is much more efficient and less cognitively demanding than reading through a researcher's entire publication history.'' Furthermore, $P7$ observed that engaging with these digital personas helped differentiate candidates who shared nearly identical task match scores in the visual chart.

Participants confirmed that the \textit{Terminology Comparison} and \textit{Conversation Rehearsal} successfully \textbf{prepared them for real-world communication} (Q7: 4.67/5). Probing the simulation with controversial questions regarding data sharing, $P6$ found that the system accurately reflected a candidate's conservative stance, grounded in privacy concerns from their prior publications. $P6$ deemed this capability highly valuable for navigating sensitive topics during formal networking. $P7$ reported that simulations offered tailored advice to avoid outdated methods.

\subsubsection{System Usability and Adoption}
{\name} received high ratings for its \textbf{learnability} and \textbf{usability} (Q8: 4.75/5, Q9: 4.92/5). $P3$ found the interface intuitive to navigate. $P5$ noted that the visual encodings, specifically the \textit{Research Timeline} and \textit{Researcher Comparison}, proved accessible following the initial tutorial. Highlighting the rich interactive design, $P1$ observed that the seamless coordination across multiple visual modules significantly reduced context switching. This fluid transition helped users focus on academic evaluation.

Participants expressed a strong \textbf{reuse intention} (Q10: 4.83/5). Multiple users advocated for deploying {\name} on public academic platforms. $P3$ described the design as ``incredibly impressive'' and confirmed the intention to adopt the tool for future projects. Furthermore, $P8$ recognized the broader application potential of the system and inquired about future release plans, underscoring the practical utility of the visual analytics workflow in real-world collaborative scenarios.

\subsection{Suggestions and Future Directions}
Participants proposed several directions for future system iterations to further support interdisciplinary collaboration.

\textbf{Broadening Collaboration Context.} Real-world academic partnerships involve constraints beyond technical matching. $P1$ suggested that visualizing collaboration motives, such as rapid algorithmic development versus long-term domain impact, would help identify partners with aligned goals. Additionally, $P3$ noted that incorporating institutional affiliations could assist users in navigating policy restrictions and geographical preferences. $P8$ proposed mapping academic kinship networks through shared co-authors to reveal introduction pathways.

\textbf{Enabling Proactive Coordination in Group Rehearsals.} Evaluating team dynamics prior to formal contact remains challenging. To address this, $P9$ and $P10$ suggested expanding the \textit{Conversation Rehearsal} to support multi-party discussions with several selected researchers simultaneously \cite{pan2025agentcoord}. Within this simulated group environment, $P10$ proposed that the AI assume a proactive leadership role. The system could automatically decompose the user's Research Goal into a structured framework and actively drive the dialogue among the simulated researchers \cite{feng2026orchmas}. Additionally, the AI could strategically prompt the user for input at critical junctures to advance the conversation. This proactive coordination would alleviate the user's burden of orchestrating complex cross-disciplinary discussions from scratch.

\textbf{Modeling Academic Schools of Thought and Theoretical Conflicts.} $P6$ identified the need to model theoretical alignments and academic schools of thought. Scientific communities often contain distinct theoretical camps with conflicting foundational assumptions. Future systems could address this by mining semantic content and co-authorship networks to cluster scholars based on their theoretical inclinations. Visually encoding these distinct schools of thought would enable users to navigate theoretical conflicts or intentionally foster diverse intellectual debates before formal engagement.

\section{Component-Level Evaluation}
We further evaluate the effectiveness of three key components in {\name}: Goal Translation (Section~\ref{sec:goal-decomposition}), Task Match Scoring (Section~\ref{sec:researcher-analysis}), and Terminology Comparison (Section~\ref{sec:terminology-analysis}). Since no benchmark directly fits this evaluation setting, six expert users from the user study ($P2$, $P5$, $P6$, $P7$, $P9$, and $P11$) participated in a separate follow-up component-level evaluation, rating multiple outputs generated from their own open-ended system use.

\textbf{Evaluation Design.} We use 5-point Likert scales, with measures defined in Table~\ref{tab:component-validation-measures}, to assess whether component outputs appropriately capture users' goals (R1), support task-relevant researcher discovery (R2), and reveal distinctive terminology patterns across domains (R3). For task match scoring, we compare Time-aware Task Match Scoring with two baselines: Similarity-only and Publication-count, which rank candidates by semantic similarity and task-relevant publication counts, respectively. The three ranking methods are anonymized and presented to users in randomized order.

\begin{table}[t]
  \centering
  \caption{Measures used in the expert-user component-level evaluation.}
  \label{tab:component-validation-measures}
  \small
  \renewcommand{\arraystretch}{1.2}
  \resizebox{\columnwidth}{!}{
  \begin{tabular}{ll}
    \toprule
    \textbf{Measure} & \textbf{Definition} \\
    \midrule
    \rowcolor[gray]{0.95}
    \multicolumn{2}{l}{\textbf{Goal Translation}} \\
    Faithfulness (Faith) & Preserves the user's intended collaboration goal. \\
    Actionability (Act) & Yields concrete tasks for targeted search and evaluation. \\
    Coverage (Cov) & Covers the key needs implied by the goal. \\
    \midrule
    \rowcolor[gray]{0.95}
    \multicolumn{2}{l}{\textbf{Task Match Scoring}} \\
    Task Fit & Aligns ranked candidates with the target task. \\
    Expertise Recency (Recency) & Prioritizes recent task-relevant expertise. \\
    \midrule
    \rowcolor[gray]{0.95}
    \multicolumn{2}{l}{\textbf{Terminology Comparison}} \\
    Representativeness (Rep) & Captures representative domain concepts. \\
    Comparison Clarity (Clar) & Makes terminology gaps and overlaps easy to compare. \\
    \bottomrule
  \end{tabular}
  }
\end{table}

\begin{figure}[t]
  \centering
  \includegraphics[width=\linewidth]{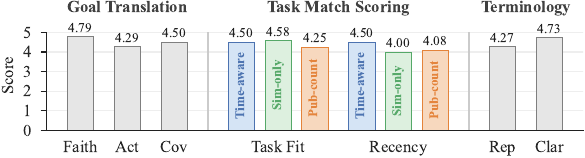}
  \caption{Mean expert-user ratings for the evaluated goal translation, task match scoring, and terminology comparison outputs.}
  \label{fig:component}
\end{figure}

\textbf{Findings.} \autoref{fig:component} summarizes the rating results across the three evaluated outputs. Goal translation received positive ratings for faithfulness (4.79), actionability (4.29), and coverage (4.50). $P7$ noted that for an early-stage, open-ended goal, the generated tasks faithfully and comprehensively captured the intent, but could further specify concrete next steps. For task match scoring, Time-aware maintained comparable task fit to Similarity-only (4.50 vs. 4.58) while improving recency (4.50 vs. 4.00). Terminology comparison received ratings of 4.27 for representativeness and 4.73 for clarity, suggesting its value for supporting users' understanding of terminology across domains.

\section{Discussion}
We discuss design lessons for LLM-augmented visual analytics, {\name}'s potential as a bidirectional collaborative ecosystem, workflow generalizability, and limitations of our cross-sectional evaluation.

\subsection{Design Lessons for LLM-Augmented Visual Analytics}
DeepConnect illustrates how visualization and LLM-based interaction jointly support interdisciplinary collaborator discovery. At the macro level, visualizations organize task, paper, and temporal evidence for side-by-side candidate comparison, helping users narrow the search space without relying solely on textual dialogue over fragmented evidence. At the micro level, LLM-based interaction enables flexible probing of feasibility, knowledge boundaries, and communication strategies for a project idea. Together, they provide a grounded, adaptive workflow for exploratory assessment.

These strengths suggest tighter coupling between visual and LLM-based interactions. For example, LLMs could explain unfamiliar concepts in terminology views, while visualizations could expose evidence behind LLM-based dialogue, reveal agent behavior, and diagnose retrieval failures or hallucination risks \cite{lu2025agentlens, tian2026ragexplorer, wang2025xgraphrag}. Future systems should make visual and language-based interactions mutually informative for transparent, coherent analysis.

\subsection{Transitioning to a Collaborative Ecosystem}
Currently, the system uses academic publications as proxies for human expertise. A broader platform could support direct exchange of methodological and physical resources. Future versions could also allow researchers to publish collaboration intents and available capabilities, turning the current egocentric workflow into bidirectional matchmaking. By integrating user-uploaded assets and domain-specific repositories, the system could map shared scientific assets, from datasets and models to specialized equipment and compounds. Together, these extensions would move the tool from personnel discovery toward a broader ecosystem for allocating intellectual and tangible scientific resources.

\subsection{System Generalizability}
Although the current implementation uses OpenAlex, the workflow is data- and domain-agnostic. Goal translation, semantic retrieval, and visual cognitive alignment could adapt to other structured and unstructured data, including preprint repositories, patent databases, and grant records. Such sources would capture early-stage research trends before peer review, giving users a more immediate view of emerging interdisciplinary opportunities. They could also reveal emerging expertise before it appears in mature citation or publication histories.

\subsection{Longitudinal Evaluation Limitations}
A primary limitation is the cross-sectional evaluation. Although participant feedback suggests that \textit{DeepConnect} may reduce immediate cognitive load during exploration and communication preparation, interdisciplinary partnerships develop over time. Our framework focuses on the pre-collaboration phase and omits long-term team viability and relational dynamics. Future work should conduct longitudinal, in-the-wild deployments across institutions, connecting early exploration behavior with subsequent collaboration milestones. Tracking follow-up contacts and joint outputs would clarify how early exploration shapes sustained collaboration over time.

\section{Conclusion}
Initiating interdisciplinary collaboration requires goal translation, candidate evaluation, and semantic common ground. We present {\name}, an LLM-augmented visual analytics system that turns broad goals into actionable tasks, supports evidence-based \textit{Researcher Exploration} with visual evidence of trajectories, fit, and complementarity, and bridges cognitive gaps through \textit{Terminology Comparison} and publication-grounded \textit{Conversation Rehearsal}. Across two case studies, a 12-researcher user study, and a component-level evaluation, the findings suggest reduced cognitive load, context-aware assessment, and more communicable early collaboration ideas.

\acknowledgments{%
This research is supported by the National Research Foundation, Singapore and Infocomm Media Development Authority under its Trust Tech Funding Initiative. Any opinions, findings and conclusions or recommendations expressed in this material are those of the authors and do not reflect the views of National Research Foundation, Singapore and Infocomm Media Development Authority.%
}

\bibliographystyle{style/abbrv-doi-hyperref}

\bibliography{reference}

\end{document}